\documentclass[pra,aps,superscriptaddress,showpacs,twocolumn]{revtex4}

\usepackage{epsfig}
\usepackage{bm}                 
\usepackage{amsmath}
\usepackage{amssymb}

\begin{document}

\title{Properties of multi-partite dark states}

\author{Pieter Kok}\email{pieter.kok@jpl.nasa.gov} 
\affiliation{Informatics, Bangor University, Bangor LL57 1UT, UK;}
\affiliation{Jet Propulsion Laboratory, California Institute of Technology,
         Mail Stop 126-347, 4800 Oak Grove Drive, Pasadena, California
         91109-8099;}

\author{Kae Nemoto}\email{nemoto@informatics.bangor.ac.uk} 
\affiliation{Informatics, Bangor University, Bangor LL57 1UT, UK;}

\author{William J.\ Munro}\email{bill_munro@hp.com} 
\affiliation{Hewlett-Packard Laboratories, Filton Road, Stoke Gifford,
         Bristol, BS34 SQ2, UK.}

\date{\today}
\begin{abstract}
 We investigate and define dark and semi-dark states for multiple
 qudit systems. For two-level systems, semi-dark and dark states are
 equivalent. We show that the semi-dark states are equivalent to the
 singlet states of the rotation group. They exist for many
 multiple qudit systems, whereas dark states are quite rare. We then
 show that when a dark state is collapsed onto another dark state of fewer
 parties, the resulting state is again dark. Furthermore, one can use two
 orthogonal multi-qudit dark states to construct a decoherence-free qudit.
\end{abstract}
\maketitle

Quantum computation and communication relies in part on the controlled
evolution of quantum systems \cite{divincenzo00}. Any uncontrolled
influences (for example from the environment) will generally cause
errors in the outcome of the computation. This effect is called
decoherence. There are two ways in which decoherence can be
overcome. Firstly, we can design our algorithms in such a way that
errors can be traced. This allows us to perform so-called quantum
error correction \cite{shor95,steane96}. Secondly, we can prevent
decoherence from happening by making the quantum system insensitive to
the environment. Basically we aim to retreat into a quiet part of
Hilbert space where the effects of decoherence are small. In both
approaches quantum information is protected against decoherence by
encoding it into entangled superpositions of multiple-qubit states
with special symmetry properties. It is currently believed that for
scalable quantum computation and communication both techniques could be
required simultaneously. One would encode qubits (or
qudits) in near-decoherence-free states. The resulting (small) errors
can then be actively corrected using quantum error correction
techniques. 

The prevention of decoherence for certain quantum states has already
been achieved in several optical cavities
\cite{aspect88,beige00a,beige00b,kis01}. Such states are called {\em
  dark states}, and they are the eigenstates of the interaction
Hamiltonian with eigenvalue zero. This ensures that these states do
not evolve in time, a property which is also exploited in, for
example, quantum clock synchronisation \cite{jozsa00}. More generally,
if we have a number of states which are invariant under a specific
class of unitary transformations, these states are said to span a
so-called decoherence-free subspace
\cite{palma96,duan97,duan98,zanardi97a,zanardi97b,lidar98}. A
decoherence-free qubit (or qudit) can then be encoded from such
states. 

In this paper we study dark states in terms of classes of unitary
transformations that leave the state unchanged. We will consider $N$
parties which all undergo the same unitary transformation $U$. This
implies that all parties have the same dimensionality. An $N$-party
pure dark state $|\Psi_N\rangle$ is then defined as 
\begin{equation}\label{dark}
 U^{\otimes N} |\Psi_N\rangle \equiv U\otimes U \otimes \ldots \otimes
 U \;\;|\Psi_N\rangle=|\Psi_N\rangle\; .
\end{equation}
A special class of dark states is given by the states which are
invariant under {\em any} arbitrary transformation $U^{\otimes N}$. An
example of such a state (for $N=2$) is the anti-symmetric pure Bell state
$|\Psi^-\rangle = (|0,1\rangle - |1,0\rangle) /\sqrt{2}$. It is well
known that $U\otimes U|\Psi^-\rangle = |\Psi^-\rangle$ for any $U$. 
Hereafter, we ignore the global phase change by the unitary operation
in Eq.\ (\ref{dark}). We define a subclass of states (called {\it
  semidark states}) which remain unchanged under only $SU(2)$ unitary
transformations.

Dark states have been classified for bipartite systems by Werner,
including mixed as well as pure states \cite{werner89}. Given any
unitary transformation $U$ of a $d$-dimensional system, a dark state
$\rho$ must satisfy
\begin{equation}
 (U \otimes U) \;\rho\; (U^{\dagger} \otimes U^{\dagger}) = \rho\; ,
\end{equation}
Such states can be written as:
\begin{equation}\label{Werner}
 \rho = \alpha I + \beta V\; ,
\end{equation}
where $I$ is the identity operator and $V$ the flip operator: $V
|\phi\rangle |\psi\rangle = |\psi\rangle |\phi\rangle$. When we
require $\rho$ to be pure, it is easily verified that the only
two-level (spin-$\frac{1}{2}$) bipartite dark state is the
anti-symmetric Bell state $|\Psi^-\rangle$. The set of mixed
dark states is larger than the set of pure dark states. In this
article we will primarily focus on pure dark states of $N$ parties,
but we will return to mixed dark states briefly later.

We have noted earlier that the anti-symmetric Bell state $|\Psi^-\rangle$
is a pure two-qubit dark state for any unitary transformation of the form
$U\otimes U$, which is also known as a {\em singlet} state. More
generally, a singlet state $|\Psi\rangle$ for $N$ $d$-level systems is defined
by
\begin{equation}
 J_{\pm} |\Psi\rangle = 0\; ,
\end{equation}
where $J_{\pm}$ are defined as
\begin{equation} \label{J}
 J_{\pm} \equiv J_{\pm}^{(1)}+J_{\pm}^{(2)}+\cdots +J_{\pm}^{(N)}\; .
\end{equation}
The operators $J_{\pm}^{(j)}$ are the $SU(2)$ ladder operator for the
angular momentum for each $d$-level subsystem $j$ with $j=1,2,\ldots N$.
These operators generate the irreducible representations of the
rotation (covering) group. The third generator is given by $J_0 =
\frac{1}{2} [J_+,J_-]$. The three operators form a closed Lie-algebra.

By definition, the most general unitary transformation of a $d$-level system
is an element of the group $SU(d)$. For convenience we will introduce the
following notation: a unitary transformation $U$ acting on a $d$-level system
is an element of $SU(d)$, whereas a unitary transformation $R$ acting on a
$d$-level system is an element of $SU(2)$. We can now prove our first theorem.

\medskip

{\bf Theorem 1:} An $N$-party pure quantum state is dark under $SU(2)$
transformations if and only if it is a singlet state:
\begin{equation}
 J_{\pm} |\Psi_N\rangle = 0 \quad\Longleftrightarrow\quad R^{\otimes N}
 |\Psi_N\rangle = |\Psi_N\rangle\; .
\end{equation}

\medskip

Any $N$-party $d$-level state $|\Psi\rangle$ that is invariant under
the transformation $U^{\otimes N}$ is called dark. By contrast, when
$|\Psi\rangle$ is only invariant under $R^{\otimes N}$, we will call
it {\em semi-dark}.

\medskip

{\bf Proof:} To begin, note that in the theory of angular momentum
$J_{\pm}|\Psi_N\rangle=0$ implies $j=0$ and $m=0$. This means that when
$J_{\pm} |\Psi_N\rangle = 0$, this automatically sets $J_0 |\Psi\rangle = 0$,
with $J_0$ the third generator of $SU(2)$.

To begin we will first prove necessity ($\Rightarrow$): since it is
sufficient to show that the theorem holds for infinitesimal rotations
over angles $\beta_k$ (all $SU(2)$ group elements are continuously
connected to the identity), we assume that $\beta_k \ll 1$. Note that
$R^{\otimes N}$ then can be written as $1+i\sum_k\beta_k J_k +
O(\beta^2_k)$ (with $k\in\{+,-,0\}$ and $J_k = J_k^{(1)} + J_k^{(2)} +
\ldots + J_k^{(N)}$):
\begin{equation}\label{rotation}
 R^{\otimes N} |\Psi_N\rangle = \left( 1 + \sum_{k\in\{+,-,0\}
   }i\beta_k J_k \right) |\Psi_N\rangle\; .
\end{equation}
Since the values of $\beta$ are equal for all $R$'s, it is immediately clear
that $J_{\pm} |\Psi_N\rangle = 0 \Rightarrow R^{\otimes N}|\Psi_N\rangle =
|\Psi_N\rangle$. This proves the necessity of being an $SU(2)$ singlet.

Now, we prove the sufficiency ($\Leftarrow$): by writing $R^{\otimes N}$ in
its infinitesimal form [see Eq.\ (\ref{rotation})] we obtain
\begin{equation}
 R^{\otimes N}|\Psi_N\rangle = |\Psi_N\rangle ~\Longrightarrow
 \sum_{k\in\{+,-,0\} } \beta_k J_k |\Psi_N\rangle = 0\; .
\end{equation}
When we define $J_+ |\Psi_N\rangle \equiv |\phi_N^+\rangle$,
$J_- |\Psi_N\rangle \equiv |\phi_N^-\rangle$ and  $J_0 |\Psi_N\rangle \equiv
|\phi_N^0\rangle$, we obtain the expression
\begin{equation}
 \beta_+ |\phi_N^+\rangle + \beta_- |\phi_N^-\rangle + \beta_0
 |\phi_N^0\rangle = 0\; .
\end{equation}
Since $\beta_+$, $\beta_-$ and $\beta_0$ are linearly independent parameters,
this implies that $J_k |\Psi_N\rangle$ must vanish for every $k$:
\begin{equation}
 J_{\pm} |\Psi_N\rangle = 0 ~\mbox{and}~ J_0 |\Psi_N\rangle = 0\; .
\end{equation}
We therefore have $R^{\otimes N}|\Psi_N\rangle = |\Psi_N\rangle \Rightarrow
J_{\pm} |\phi_N\rangle = 0$. This completes the proof.\hfill$\square$

\medskip

We now extend our analysis to dark states in $N$-party $d$-level
systems and consider whether dark states exist, that is, whether there
are states that satisfy Eq. (\ref{dark}). It is convenient to employ
ladder operators for the $SU(d)$ operation. There are $2d(d-1)$ ladder
operators for $SU(d)$: $d(d-1)$ operators for each of raising and
lowering.  An $SU(d)$ ladder operator $J_{\pm  (hj)}$ for each $d$-level
system can be considered as an operation on the subsystem of $h$ and
$j$ levels, where $h\neq j$ and $1\leq h,j\leq d$.
This leads to our next theorem.

\medskip

{\bf Theorem 2:} Any pure $N$-party, $d$-level quantum state
$|\Psi\rangle$ is dark if and only if all possible $SU(d)$ ladder
operators map $|\Psi\rangle$ onto zero:
\begin{equation} \label{theorem12}
 J_{\pm (h,j)} |\Psi_N\rangle = 0 ~\Longleftrightarrow~ U^{\otimes N}
 |\Psi_N\rangle = |\Psi_N\rangle\; ,
\end{equation}
for all $1\leq h,j\leq d$ .

\medskip

{\bf Proof:} To begin our proof, we use the fact that any $SU(d)$
matrix, that is, a general unitary transformation of an $d$-level system, can
be decomposed as $SU(d-1)_{(2,\ldots,d)}  SU(2)_{(1,2)} 
SU(d-1)_{(2,\ldots,d)} $, where the superscript denotes the levels the
group elements act on \cite{Rowe99}. Repeating this decomposition for
every $SU(d')$ with $d'>2$, the matrix can be expressed in terms of
$SU(2)_{j,j+1}$, where $1\leq j\leq n-1$. Hence  Theorem 1 guarantees
$J_{\pm (j,j+1)} |\Psi_N\rangle = 0 \Leftrightarrow \quad
SU(2)_{(j,j+1)}^{\otimes N} |\Psi_N\rangle = |\Psi_N\rangle$. If the
state $|\Psi_N\rangle$ satisfies the right hand side of
(\ref{theorem12}), then the above relation must hold for any $j$. 

We first prove necessity ($\Rightarrow$). It is clear from the
preceding discussion that the above condition for $SU(2)_{j,j+1}$ is a
necessary condition of the left-hand side of (\ref{theorem12}), so the
right hand side of (\ref{theorem12}) always holds. 

Now, we prove sufficiency ($\Leftarrow$). The above relation gives
$J_{\pm (j,j+1)} |\Psi_N\rangle = 0$ for any $j$. This implies the
relation $J_{\pm (j,j+1)}J_{\pm (j+1,j+2)}=J_{\pm (j,j+2)}$, hence
obtains the left hand side of (\ref{theorem12}). This completes the
proof. \hfill$\square$
 
\medskip

This proof can then be used to show that two $d$-level ($d>2$)
systems have no dark states. This is shown in the following corollary:

\medskip

{\bf Corollary 1:} There are no pure $d$-level, bi-partite dark states
(for $d>2$). 

\medskip

{\bf Proof:} Let $|a^{(1)}, a^{(2)}\rangle$ be a state of a $d$-level,
bi-partite system, where $a$ is an integer for an odd number of $d$
or a half-integer for an even number of $d$ in $-(d-1)/2\leq
a\leq (d-1)/2$.  The suffix of $a$ is to distinguish the qudits. It is
necessary for a dark state to satisfy the conditions for semi-dark
states, so that we can require $a^{(1)}+a^{(2)}=0$. In other words,
any candidate bi-partite dark state $|\psi\rangle$ must be some
superposition of states $|m,-m\rangle$. Furthermore, this must remain
true after $SU(d)$ bit-flip operations. However, for $d>2$ there
exists at least one bit-flip operation that maps the state $|a^{(1)},
a^{(2)}\rangle$ onto $|\bar{a}^{(1)},\bar{a}^{(2)}\rangle$ with $\bar{a}^{(1)} +
\bar{a}^{(2)} \neq 0$. This means that there is no state that can be a
component of a dark state for $d>2$, hence there are no pure
$d$-level, bi-partite dark states. \hfill $\square$

This result was first proved by Werner \cite{werner89}.

\medskip

As an example to illustrate this corollary, consider the two-qutrit
(spin one) state 
\begin{equation}
 |\phi\rangle = \frac{1}{\sqrt{3}} \left(|1,-1\rangle + |-1,1\rangle -
 |0,0\rangle \right)\; ,
\end{equation}
It is straightforward to show that this state is not dark, even though
it has $j=0$ and $m=0$.  A bit-flip operation on the levels of
$|0\rangle$ and $|1\rangle$, remaining the state $|-1\rangle$
unchanged, maps the state to another state $(|0,-1\rangle + |-1,0\rangle - 
|1,1\rangle)/\sqrt{3}$.  Hence there are some $SU(d)$ operators exist to
change the state $|\phi\rangle$, while $SU(2)$ operators preserve the
state unchanged.  An extention of the corollary above to the
$N$-partite case is now straightforward.

\medskip

{\bf Corollary 2:} There are no dark states in $N$-party $d$-level
systems if $N < d$.

\medskip

{\bf Proof:} Let $|a^{(1)}, \ldots, a^{(N)}\rangle$ be a state of an
$N$-party $d$-level system, where $a$ is integer or half-integer
in $-(d-1)/2\leq a \leq (d-1)/2$ depending on its parity. It is
necessary for a dark state to satisfy a condition for semi-dark
states, which is $\sum^N_{j=1} a^{(j)}=0$. We use this condition to
restrict states to be analysed as we have seen in Corollary 1. The
action of bit-flip operators maps the set $\{a^{(j)}\}$ to another
$\{\bar{a}^{(j)}\}$. For the case of $N<d$, there is at least one bit-flip
operation which maps elements of $a^{(j)}$ to other elements not in
the original set of $a^{(j)}$, i.e. $\bar{a}^{(j)}\not\in \{a^{(j)}\}$. It
is obvious that as there are no changes in the other elements by this
mapping, the new sequence of $\bar{a}^{(j)}$ gives $\sum^N_{j=1}
\bar{a}^{(j)}\neq 0$. This directly leads to no dark states in $N$-party,
$d$-level systems $(N<d)$. \hfill$\square$

\medskip

The results from these corollaries prompt us to the following
question: given that no $N$-partite dark states exist for $d$-level
systems if $N<d$, do there exist $d$-partite, $d$-level dark states?
It turns out that the answer to this question is yes, which we will
prove by explicit construction.

\medskip

{\bf Theorem 3} The smallest system of qudits in a dark state is a
$d$-party $d$-level system.

\medskip

By virtue of Corollary 2, we only have to show that $d$-partite dark
qudit states exist. However, before we commence with the proof we
consider two examples for $d=3$ and $d=4$ (without proof).

The most general unitary transformation of a qutrit is given by an
$SU(3)$ transformation. Therefore, a system consisting of three qutrits
has a true dark state under $SU(3)^{\otimes 3}$. We can make the
following construction for such a dark state with a normalisation
factor ${\cal N}$:
\begin{eqnarray}
 |\Psi_3\rangle &=& {\cal N}\left\{|1,0,-1\rangle-|1,-1,0\rangle\right. \nonumber\\
 && +|0,-1,1\rangle -|0,1,-1\rangle \nonumber\\
 && +\left. |-1,1,0\rangle -|-1,0,1\rangle\right\},\nonumber\\
 &\equiv& {\cal N}P_{all}[|1,0,-1\rangle].
\end{eqnarray}
Here the operator $P_{all}$ is defined as the sum of all possible
states generated by repeating pair-wise permutations with a relative sign
flip. Note the absence of $|0,0,0\rangle$ in this superposition which
is also a $j=0$, $m=0$ state.

Using the same technique, we can construct the dark state for
four-party four-level systems:
\begin{eqnarray}
 |\Psi_4\rangle &=&
{\cal N} S^{(1,2)}S^{(3,4)}S^{(1,4)}
 \Big(|3/2,1/2,-1/2,-3/2\rangle \nonumber\\
 && + |-1/2,3/2,1/2,-3/2\rangle \nonumber\\
 && + |1/2,-1/2,3/2,-3/2\rangle\Big)\nonumber\\
 &\equiv& {\cal N}P_{all}[|3/2,1/2,-1/2,-3/2\rangle],
\end{eqnarray}
where $S^{(j,k)}(*)$ is defined as the partial $SU(2)$ singlet
operator, which acts on the $j$-th and $k$-th qudits to generate a
single state for this subsystem with respect to the given values of
$a^{(j)}$ and $a^{(k)}$.  For instance,
\begin{equation}
  S^{(1,3)}(|\alpha,a^{(2)},\beta\rangle)
  ~\longrightarrow~
  |\alpha,a^{(2)},\beta\rangle - |\beta,a^{(2)},\alpha\rangle.
\end{equation}
The use of repeated $S^{(j,k)}$'s to generate dark states is closely related
to the decomposition of $SU(d)$, as a result, the state is tolerant to
$SU(2)$ operations on any subsystems.  A simple extension of these dark
states to the general $d$-party $d$-level dark states suggests
\begin{equation} \label{dddark}
  |\Psi_d\rangle= {\cal N} P_{all}[|-(d-1)/2,\ldots,(d-1)/2 \; \rangle].
\end{equation}
This ansatz allows us to prove Theorem 3, including the above examples.

\medskip

{\bf Proof of Theorem 3:} We prove that the state (\ref{dddark}) is
dark by showing that $J_{\pm (a_j,a_k)}|\Psi_d\rangle=0$ for an
arbitrary pair $(j,k)$ with $j\neq k$. We label each level in the
qudit as $a_s$ with $1\leq s\leq d$.
As the state in Eq.\ (\ref{dddark}) includes every ordering of $a_s$
once and only once, the state (\ref{dddark}) is a superposition of $d!$ states.
The number of all the possible locations of a pair $(a_j,a_k)$ is
$d(d-1)/2$, which is equal to the number of combination to select two locations 
of $x$-th and $y$-th from $1\leq x<y \leq d$.  
For each location of the pair there are
$(d-2)!$ different combinations for the rest of the qudits.  
Therefore for a given pair $(j,k)$ the state (\ref{dddark}) can always
be re-written as
\begin{eqnarray}
|\Psi_d\rangle = &\sum_{(x,y)}& (-1)_{(x,y,j,k)}
\Big( |a_j^{(x)},a_k^{(y)}\rangle -|a_j^{(y)},a_k^{(x)}\rangle\Big)
\nonumber\\ 
&\otimes& P_{all} \Big[\; |\ldots, a_h^{(z)},\ldots\rangle_{h\neq j,k;
z\neq x,y} \; \Big],
\end{eqnarray}
where the sum is taken for all combination of $x$ and $y$, and   
$(-1)_{(x,y,j,k)}$ can be either $+1$ or $-1$ determined by the parameters, 
$(x,y,j,k)$.  From this expression of (\ref{dddark}) and the theorem 2, 
it is now clear that the action of $J_{\pm (a_j,a_k)}$ on
$|\Psi_d\rangle$ results in zero, hence the state (\ref{dddark}) is dark. 
\hfill$\square$

\medskip

At this point, it should be clear that there are no dark states for an
$N$-party $d$-level system if $d< N< 2d$, and indeed there are dark
states only if $N=md$, where $m\in {\Bbb N}$, the set of natural
numbers. For the case of $N\neq md$ we can apply the argument of
Corollary 2 to show that there are no dark states. Hence we have a
very explicit criterion for the existence of $N$-party $d$-level
systems. The method in Theorem 3 also provides an explicit recipe for
generating the dark state. To illustrate this, we re-examine
$N$-partite qubit systems.

We know that for three qubits, there are no singlet states ($m=N/d$ is
not an integer). Hence let us consider four two-level systems. The
sixteen-dimensional Hilbert space can be decomposed into ${\mathbf
  5}\oplus{\mathbf 3}\oplus {\mathbf 3}\oplus{\mathbf 3}\oplus{\mathbf
  1}\oplus{\mathbf 1}$ irreducible representations of $SU(2)$. Up to
permutation 
symmetry there are two singlet states, which can be written as
$|\Psi^-\rangle_{12} \otimes|\Psi^-\rangle_{34}$ and
$|\Psi^-\rangle_{13}\otimes |\Psi^-\rangle_{24}$. This states are
obviously dark. More generally, a linear superposition of these dark
states is also dark and this can be used to create a decoherence-free
qubit.

\medskip

{\bf Theorem 4:} Linear superpositions of two dark states are also
dark. We will prove this in two parts: for a coherent and incoherent
superposition. 

\medskip

{\bf Proof}: A {\it coherent superposition} of dark states is also
dark. To prove this, consider two dark states
$|\Psi_{N}\rangle$ and $|\Phi_{N}\rangle$. These satisfy $U^{\otimes
  N} |\Psi_N\rangle = |\Psi_N\rangle$ and $U^{\otimes N}
|\Phi_N\rangle = |\Phi_N\rangle$. Hence a coherent superposition of
these states 
\begin{eqnarray}
 U^{\otimes N} \left[ a_1 |\Psi_N\rangle +a_2 |\Phi_N\rangle \right]
 &=& a_1 \;U^{\otimes N} |\Psi_N\rangle +a_2 \;U^{\otimes N}
 |\Phi_N\rangle 
 \nonumber \\
 &=& a_1 |\Psi_N\rangle +a_2  |\Phi_N\rangle
\end{eqnarray}
is also a dark state. This proves a linear coherent superposition of
two dark state is also dark. 

We will now prove that an {\it incoherent superposition} of dark
states is also dark. Consider two dark states $\rho_1=|\Psi_N\rangle
\langle \Psi_N |$ and $\rho_2=|\Phi_N\rangle \langle \Phi_N |$. An
incoherent superposition of these dark states can be written as
$\rho=a_1 \rho_1 + a_2 \rho_2$ and hence
\begin{eqnarray}
 U^{\otimes N} \rho U^{\otimes N} &=&U^{\otimes N} \left[ a_1 \rho_1 +
 a_2 \rho_2\right] U^{\otimes N}\nonumber \\ &=& a_1 U^{\otimes N}
 \rho_1 U^{\otimes N} + a_2 U^{\otimes N}\rho_2 U^{\otimes N}\nonumber
 \\ &=&  a_1 \rho_1 + a_2 \rho_2=\rho \nonumber
\end{eqnarray}
which concludes the proof. \hfill$\square$

\medskip

The first part of this theorem is critical when one examines
decoherence-free subspaces which are formed from dark states. There
are two (unnormalised) orthogonal 4-partite qubit states: 
\begin{equation}\nonumber
  |0011\rangle + |1100\rangle + |0110\rangle + |1001\rangle -
  2|0101\rangle - 2|1010\rangle
\end{equation}
and
\begin{equation}\nonumber
  |0011\rangle + |1100\rangle - |0110\rangle - |1001\rangle
\end{equation}
Theorem 4 tells us that (coherent) superpositions of these two states
are also dark, and they therefore generate a two-dimensional
decoherence-free subspace. Since this is a two-dimensional Hilbert
space, it can be used to encode a qubit \cite{decoherence}. When there
is no interaction between the four qubits, and they share a common
environmental decoherence, then such a compound qubit suffers much
less from this form of decoherence. We call this construction a {\it
  decoherence-free qubit}. 
 
It also seems possible to encode a decoherence-free qudit in an 
analogous way to the qubit case. Here instead of 4 qubits being
necessary for the construction, $d^2$ qudits are necessary. This would
require the following conjecture to be true:

\medskip

{\bf Conjecture:} for $N = m d$ qudits, one can construct $m$
orthogonal dark states. 

\medskip

While this is true for two and four qubits, we do not have a general
proof. For systems with large $d$, this would require $d^2$ qudits all
sharing the same environment. In actual physical implementations this
will provide a practical limitation on how large $d$ can be. It does,
however, mean that error resistant computation and communication may
be possible in a commonly shared noisy environment.  

\bigskip

Our final theorem is prompted by the question how dark states behave
under wavefunction collapse. If one considers an $N$-party dark state
and projects out a $M$-party dark state what is the status of the
$N-M$ remaining state? It turns out to be dark as well.

\medskip

{\bf Theorem 5:} When the $N$-party state $|\Psi_N\rangle$ and the $M$-party
state $|\Psi_M\rangle$ are both dark (with $M<N$), then the $N-M$-party state
which results when $|\Psi_N\rangle$ is collapsed onto $|\Psi_M\rangle$ is also
dark.

\medskip

{\bf Proof:} Consider the following identities:
\begin{eqnarray}
 \langle\Psi_M|\Psi_N\rangle &=&
 \sum_i c_i\langle\Psi_M|\phi_M^i\rangle\otimes|\phi_{N-M}^i\rangle\cr
 &=& \sum_i d_i |\phi_{N-M}^i\rangle \equiv |\Psi_{N-M}\rangle\; .
\end{eqnarray}
and
\begin{eqnarray}
 \langle\Psi_M|\Psi_N\rangle &=&
 \sum_i c_i \langle\Psi_M|U^{\otimes M}|\phi_M^i\rangle 
 U^{\otimes N-M} |\phi_{N-M}^i\rangle \cr
 &=& \sum_i d_i U^{\otimes N-M}|\phi_{N-M}^i\rangle\cr &=&
 U^{\otimes N-M}|\Psi_{N-M}\rangle\; .
\end{eqnarray}
We therefore have $|\Psi_{N-M}\rangle = U^{\otimes N-M}
|\Psi_{N-M}\rangle$. This completes the proof.\hfill$\square$.

\bigskip

In this article we have studied dark states and some of their
properties. These states are critical in the formation of
decoherence-free subspaces, and thus for fault-tolerant quantum
computation. If several qudits can be placed in a common environment,
then it is possible to use multiple dark states to encode a
decoherence-free qudit. For example, in systems of four qubits, two
orthogonal dark states exist. These states can be used to encode a
decoherence-free qubit. Furthermore, we have shown that one needs at
least (a multiple of) $d$ qudits to create a dark state.

The authors would like to thank S.L.\ Braunstein for seminating
comments, questions and discusions. 
A portion of the research in this
paper was carried out at the Jet Propulsion Laboratory, California
Institute of Technology, under a contract with the National
Aeronautics and Space Administration. 
P.K.\ would like to acknowledge the National Research Council.

\end{document}